\documentclass{elsart}

\begin{document}
\begin{frontmatter}

\title{Enhanced low energy fusion rate in palladium (Pd) due to vibrational deuteron dipole-dipole interactions and associated resonant tunneling that over-cancels the Coulomb barrier between neighbouring interstitial deuteronic pair wavefunctions  }
\author{J.S.Brown}
\address{Clarendon Laboratory, 1 Parks Rd, Oxford, UK }

\begin{abstract}
It is observed that interstitial hydrogen nucleii on a metallic lattice are strongly coupled to their near neighbours by the unscreened electromagnetic field mediating transitions between low-lying states. It is shown that the dominant interaction is of dipole-dipole character. By means of numerical calculations based upon published data, it is then shown that in stoichiometric PdD, in which essentially all interstitial sites are occupied by a deuteron, certain specific superpositions of many-site product states exist that are lower in energy than the single-site ground state, suggesting the existence of a new low temperature phase. Finally, the modified behaviour of the two-particle wavefunction at small separations is investigated and prelimary results suggesting a radical narrowing of the effective Coulomb barrier are presented.
\end{abstract}

\end{frontmatter}

\section{Introduction}
\label{intro} 
Several metallic elements, notably palladium, vanadium, niobium and nickel, can reversibly absorb hydrogen up to the point of stoichiometry, in which every available interstitial site - of octahedral (O) or tetrahedral (T) symmetry - is occupied by a hydrogen nucleus. A single hydrogen nucleus in such an environment exhibits a spectrum of singlet, doublet and triplet state representations of the local point symmetry group. The ground state is invariably  a singlet with even $[+++]$ parity along each of the symmetry axes. The next level is, in an fcc lattice, a triplet of states with parities $[-++]$, $[+-+]$, $[++-]$ and an excitation energy of the order of 60 meV. The dipole moment between the ground state singlet and the first excited triplet is typically of the order of $0.2\AA e$. Since the electronic Fermi gas couples weakly to the electromagnetic field quanta in this part of the spectrum, such a dipole moment gives rise to an essentially unscreened resonant dipole-dipole interaction (RDDI) between nearest neighbours \cite{Kur96}. Simple geometrical considerations reveal this to be of the order of 20 meV per pair. Not only is this typically several times larger than screened static Coulomb interaction, it is also manifestly an appreciable fraction of the (on-diagonal) excitation energy of the dipole itself. Since in the quasi-stoichiometric loading regime each hydrogen has several nearest neighbours, it has previously been speculated {\cite{Bro06} that there exist many-site states for which the total collective effect of the interaction is a multiple of the pair interaction. This paper sets out to answer the, in our opinion, intriguing question as to whether there exists any such collective state of quantum-entangled dipoles whose total energy is lower than the simple product of ground states, and to obtain an upper bound on the lowest possible energy of such an ensemble of coupled oscillators.
\section{Model}
The single particle states $\psi_n$ are the solutions of
\begin{equation}
\label{Schro1}
H(r)\psi_n({\bf r}) = \left[ {-\hbar^2 \over 2M_H} \nabla^2 + V({\bf r}) \right] \psi_n({\bf r}) = \epsilon_n \psi_n({\bf r})
\end{equation}
--where V(r) is the periodic potential experienced by an infinitely heavy positive charge with fixed metal core positions. The reader is directed to \cite{Kri94} for a detailed discussion of the derivation of this potential using the DFT procedure. In view of the relatively large mass $M_H$ of hydrogen nucleii, the lowest energy solutions will generally be well-localised about local minima in V. These minima will coincide with sites of octahedral or tetrahedral symmetry in cubic lattices and hence the levels $\epsilon_n$ are an assortment of singlets, doublets and triplet representations of the cubic point symmetry groups. The static (zero frequency) components of the potential disturbance due to the hydrogen nucleus is subject to a screening law of the approximate (Thomas-Fermi) form
\begin{equation}
\label{TF}
V_{HH}(r) =  e^2{e^{ -Kr } \over r}
\end{equation}
where $K$ is proportional to the DOS at the Fermi surface.
\\
K is typically much greater than a reciprocal lattice vector. The static Coulomb H-H interaction between nearest neighbours is consequently small, typically not more than a few meV, and essentially state-independent. By contrast, the attenuation of the electromagnetic field due to a transition between levels is negligible over the dimensions of a lattice cell.
\\
Since the interparticle interaction is so strongly frequency dependent, the full Hamiltonian cannot be written in closed analytical form. 
However, matrix elements between pairs of two-site states are simply given by:
\begin{equation}
\label{hif}
H_{i_1,j_1;i_2,j_2} = 
e^2\langle j_1,j_2 | 
{e^{-K  |{\bf r}_1-{\bf r}_2 - {\bf R}| \delta(\epsilon_{i_1} - \epsilon_{j_1}) 
\delta(\epsilon_{i_2} - \epsilon_{j_2})} 
\over |{\bf r}_1-{\bf r}_2 - {\bf R}| } 
| i_1,i_2 \rangle
\end{equation}
where $\bf R$ is the intersite displacement vector. The $\delta(\epsilon_{i} - \epsilon_{f})$ factors express the fact that only transitions between different levels give rise to an unscreened interaction.
\\
If just the lowest order term in the multipole expansion of the Coulomb operator $1 \over |{\bf r_1} - {\bf r_2} - {\bf R}|$ is retained, there is no need for double integration over both volumes. In this approximation, (\ref{hif}) reduces to the familiar expression for a dipole-dipole interaction:
\begin{equation}
\label{dipdip}
H_{i_1,j_1;i_2,j_2} \approx {e^2 \over R^3} \langle j_1 |{\bf r_1}|i_1\rangle \cdot \langle j_2 |{\bf r_2}|i_2\rangle
- {3e^2 \over R^5} \langle j_1 |{\bf R \cdot r_1}|i_1\rangle \langle j_2 |{\bf R \cdot r_2}|i_2\rangle
\end{equation}
If a classical dipole is located at every interstitial O-site in an fcc lattice, the interaction energy is lowest with the following orientations over a constant-z plaquette:
\begin{equation}
\label{xyplane}
\left[
\begin{array}{ccccccccc}
\leftarrow & \circ & \rightarrow &\circ & \leftarrow & \circ & \rightarrow & \circ & \leftarrow
\\
\circ& \uparrow &\circ& \downarrow &\circ& \uparrow &\circ& \downarrow & \circ
\\
\rightarrow &\circ& \leftarrow &\circ& \rightarrow &\circ& \leftarrow & \circ & \rightarrow
\\
\circ& \downarrow &\circ& \uparrow &\circ& \downarrow &\circ& \uparrow & \circ
\\
\leftarrow &\circ& \rightarrow &\circ& \leftarrow &\circ& \rightarrow & \circ & \leftarrow
\end{array}
\right]
\end{equation}
- where the open circles represent the sites of the metal cores at locations [{\bf 100}], [{\bf 300}],[{\bf 111}] etc. 
\\
It can be shown that there is zero net interaction between parallel layers when each layer has such an arrangement.
Guided by this classical analogue, we will limit our search for minimum energy states to those constructed from:
\\[0.3cm]
$[+++]$ parity states at all O-sites in a $z=0$ plaquette
\\
$[-++]$ parity states at O-sites of even $y$ and 
\\
$[+-+]$ states at O-sites of odd $y$. 
\\
$[+++]$ ground states at all sites external to the plaquette
\\[0.3cm]
For the rest of this paper we will use the shorthand  $| s, n \rangle$ to denote the $n$th state of [$+++$] parity, $| p_x, n \rangle$ to denote the $n$th state of [$-++$] parity and $| p_y, n \rangle$ to denote the $n$th state of [$+-+$] parity. The singlet ground state is accordingly written as $|s,0\rangle$. If the site location needs to be made explicit, we will append this in bold type thus: $|s,0, {\bf 110} \rangle$. 
For clarification, we reproduce below an example of a pair of five-O-site states that are degenerate in zeroeth order and that are linked by the dipole-dipole interaction of (\ref{dipdip}):
\begin{eqnarray}
\left[
\begin{array}{ccc}
|s,0 \rangle & \circ & |s,0 \rangle
\\
\circ & |p_y,0 \rangle & \circ
\\
|s,0 \rangle & \circ & |s,0 \rangle
\end{array}
\right],
\left[
\begin{array}{ccc}
|s,0 \rangle & \circ & |s,0 \rangle
\\
\circ & |s,0 \rangle & \circ
\\
|p_x,0 \rangle & \circ & |s,0 \rangle
\end{array}
\right]
\nonumber
\end{eqnarray}
or equivalently:
\begin{eqnarray}
\label{sp5}
|s,0,{\bf 000} \rangle \otimes |s,0, {\bf 020} \rangle \otimes |p_y,0, {\bf 110} \rangle
\otimes |s,0, {\bf 200} \rangle \otimes |s,0, {\bf 220} \rangle ,
\nonumber\\
|p_x,0,{\bf 000} \rangle \otimes |s,0, {\bf 020} \rangle \otimes |s,0, {\bf 110} \rangle
\otimes |s,0, {\bf 200} \rangle \otimes |s,0, {\bf 220} \rangle 
\end{eqnarray}
The Hamiltonian matrix in the subspace of the two-O-site states 
\\
$|s,0,{\bf 000} \rangle \otimes |p_y,0, {\bf 110} \rangle$ and
\\
$|p_x,0,{\bf 000} \rangle \otimes |s,0, {\bf 110} \rangle$ is, according to (\ref{dipdip}):
\begin{equation}
\label{hpair}
{\bf H} = 
\left(
\begin{array}{cc}
\epsilon_{s,0} + \epsilon_{p,0} & -{3 {\sqrt 2}e^2 d^2 \over a^3}
\\
-{3{\sqrt 2}e^2 d^2 \over a^3} & \epsilon_{s,0} + \epsilon_{p,0}
\end{array}
\right)
\end{equation}
where the dipole length $d \equiv \langle s,0 | x | p_x,0 \rangle = \langle s,0 | y | p_y,0 \rangle$
\\
The energy eigenvalues are in this case simply $\epsilon_{s,0} + \epsilon_{p,0}  \pm {3 {\sqrt 2}e^2 d^2 \over a^3}$
\\
More generally, and assuming full hydrogen occupancy and perfect lattice symmetry, the full many-site Hamiltonian matrix depends numerically upon just the lattice parameter $a$, the dipole lengths $d$ and the energies $\epsilon$ of the single-site states.
\section{Application to PdD}
Quasi-stoichiometric PdH and PdD are natural candidates for our model because the adiabatic effective potential experienced by the hydrogen nucleus has been determined from {\it ab initio} DFT calculations \cite{Kri94,Dye05} and the theoretical spectra found to agree well with IR spectroscopic measurements over a wide range of substoichiometric loading ratios. For this work we checked the results published in \cite{Kri94} by solving (\ref{Schro1}) on a real-space wedge mesh of pitch 0.03 \AA, using the published effective adiabatic Pd-H potential and boundary conditions appropriate to the desired parity. The lattice parameter, corresponding to a displacement like $[{\bf 200}]$ in our notation, is $4.07 \AA$. The lowest few eigenvectors of the very large sparse matrix were solved using the dominant-diagonal iteration method, as in \cite{Pus84}.
\\
The lowest single-site energy levels, relative to the O-site potential, were found to be:
\\[1cm]
\begin{tabular}
{|c|c|c|}
\hline
level & PdH  (meV) & PdD (meV)
\\
\hline
$\epsilon_{s,0}$ & 82 & 52
\\
$\epsilon_{p,0}$ & 151 & 95
\\
$\epsilon_{s,1}$ & 233 & 149
\\
$\epsilon_{p,1}$ & 289 & 186
\\
\hline
\end{tabular}
\\[1cm] 
- with the following dipole lengths along each of the three cartesian axes:
\\[1cm]
\\
\begin{tabular}
{|c|c|c|}
\hline
dipole & PdH (\AA) & PdD (\AA)
\\
\hline
$\langle s,0 | x | p_x,0 \rangle$ & 0.172 & 0.153
\\
$\langle s,1 | x | p_x,0 \rangle$ & 0.128 & 0.112
\\
$\langle s,0 | x | p_x,1 \rangle$ &-0.003 &-0.002
\\
$\langle s,1 | x | p_x,1 \rangle$ & 0.194 & 0.165
\\
\hline
\end{tabular}
\\[1cm] 
Double integration over both site volumes according to (\ref{hif}) yielded the following off-diagonal elements:
\\[1cm]
\begin{tabular}
{|c|c|c|}
\hline
$H_{i_1,j_1;i_2,j_2}$ & PdH (meV) & PdD (meV)
\\
\hline
$\langle s,0, {\bf 000}| \otimes \langle p_x,0, {\bf 110} 
| H | 
p_y,0, {\bf 000} \rangle \otimes |s,0, {\bf 110} \rangle$ &-27& -21
\\
$\langle s,0, {\bf 000}| \otimes \langle p_x,0, {\bf 110} 
| H | 
p_y,0, {\bf 000} \rangle \otimes |s,1, {\bf 110} \rangle$ &-20& -15
\\
$\langle s,1, {\bf 000}| \otimes \langle p_x,0, {\bf 110} 
| H | 
p_y,0, {\bf 000} \rangle \otimes |s,1, {\bf 110} \rangle$ &-15& -11
\\
$\langle s,0, {\bf 000}| \otimes \langle p_x,1, {\bf 110} 
| H | 
p_y,0, {\bf 000} \rangle \otimes |s,1, {\bf 110} \rangle$ &-30& -23
\\

$\langle s,0, {\bf 000}| \otimes \langle p_x,0, {\bf 200} 
| H | 
p_x,0, {\bf 000} \rangle \otimes |s,0, {\bf 200} \rangle$ &-13& -10
\\
$\langle s,0, {\bf 000}| \otimes \langle p_x,0, {\bf 200} 
| H | 
p_x,0, {\bf 000} \rangle \otimes |s,1, {\bf 200} \rangle$ &-9& -7 
\\
$\langle s,1, {\bf 000}| \otimes \langle p_x,0, {\bf 200} 
| H | 
p_x,0, {\bf 000} \rangle \otimes |s,1, {\bf 200} \rangle$ &-7& -5
\\
$\langle s,0, {\bf 000}| \otimes \langle p_x,1, {\bf 200} 
| H | 
p_x,0, {\bf 000} \rangle \otimes |s,1, {\bf 200} \rangle$ &-14& -11
\\

$\langle s,0, {\bf 000}| \otimes \langle p_x,0, {\bf 020} 
| H | 
p_x,0, {\bf 000} \rangle \otimes |s,0, {\bf 020} \rangle$ &6& 5
\\
$\langle s,0, {\bf 000}| \otimes \langle p_x,0, {\bf 020} 
| H | 
p_x,0, {\bf 000} \rangle \otimes |s,1, {\bf 020} \rangle$ &5& 4
\\
$\langle s,1, {\bf 000}| \otimes \langle p_x,0, {\bf 020} 
| H | 
p_x,0, {\bf 000} \rangle \otimes |s,1, {\bf 020} \rangle$ &4& 3
\\
$\langle s,0, {\bf 000}| \otimes \langle p_x,1, {\bf 020} 
| H | 
p_x,0, {\bf 000} \rangle \otimes |s,1, {\bf 020} \rangle$ &7& 5
\\

$\langle s,0, {\bf 000}| \otimes \langle p_x,0, {\bf 220} 
| H | 
p_x,0, {\bf 000} \rangle \otimes |s,0, {\bf 220} \rangle$ &-1& -1
\\
$\langle s,0, {\bf 000}| \otimes \langle p_x,0, {\bf 220} 
| H | 
p_x,0, {\bf 000} \rangle \otimes |s,1, {\bf 220} \rangle$ &-1& -1
\\
$\langle s,1, {\bf 000}| \otimes \langle p_x,0, {\bf 220} 
| H | 
p_x,0, {\bf 000} \rangle \otimes |s,1, {\bf 220} \rangle$ &-1& -1
\\
$\langle s,0, {\bf 000}| \otimes \langle p_x,1, {\bf 220} 
| H | 
p_x,0, {\bf 000} \rangle \otimes |s,1, {\bf 220} \rangle$ &0& -1
\\
\hline
\end{tabular}
\\[1cm]
\subsection{5-site states}
As an illustrative example, we will calculate the lowest 5-site energy achievable for a given number of 
$|p,0 \rangle$ states. There are just five 5-site states having 1 $|p,0 \rangle$ and 4 $|s,0 \rangle$ states, namely:
\begin{equation}
\label{five}
\left\{
\begin{array}{c}
|s,0,{\bf 000}\rangle \otimes |s,0,{\bf 020}\rangle \otimes |s,0,{\bf 110}\rangle \otimes
|s,0,{\bf 200}\rangle \otimes |p_x,0,{\bf 220}\rangle
\\
|s,0,{\bf 000}\rangle \otimes |s,0,{\bf 020}\rangle \otimes |s,0,{\bf 110}\rangle \otimes
|p_x,0,{\bf 200}\rangle \otimes |s,0,{\bf 220}\rangle
\\
|s,0,{\bf 000}\rangle \otimes |s,0,{\bf 020}\rangle \otimes |p_y,0,{\bf 110}\rangle \otimes
|s,0,{\bf 200}\rangle \otimes |s,0,{\bf 220}\rangle
\\
|s,0,{\bf 000}\rangle \otimes |p_x,0,{\bf 020}\rangle \otimes |s,0,{\bf 110}\rangle \otimes
|s,0,{\bf 200}\rangle \otimes |s,0,{\bf 220}\rangle
\\
|p_x,0,{\bf 000}\rangle \otimes |s,0,{\bf 020}\rangle \otimes |s,0,{\bf 110}\rangle \otimes
|s,0,{\bf 200}\rangle \otimes |s,0,{\bf 220}\rangle
\\
\end{array}
\right\}
\end{equation}
The corresponding Hamiltonian matrix is (in meV):
\begin{equation}
{\bf H} = 
\left(
\begin{array}{ccccc}
43 & 5 & -21 & -10 & -1
\\
5 & 43 & 21 & -1 & -10
\\
-21 & 21 & 43 & 21 & -21
\\
-10 & -1 & 21 & 43 & 5
\\
-1 & -10 & -21 & 5 & 43
\end{array}
\right)
+ 5\epsilon_{s,0}{\bf I_5}
\end{equation}
- where we have separated out the energy of the conventional ground state.
\\
The energy eigenvalues in this small subspace are 3,29,37,59 and 88 meV relative to $5\epsilon_{s,0}$. For the 80 five-site states that comprise just one $|p,0\rangle$ and four $|s,n<2\rangle$ states, the lowest two energy eigenvalues are found to be -2 and 26 meV relative to $5\epsilon_{s,0}$. It is clear from this that the $|s,1\rangle$ states make a significant contribution to the lowest energy eigenvector. 
\\
In an attempt to find a converged value for the absolute mimimum site energy, Hamiltonian matrices were constructed for a series of plaquettes of increasing size up to a limit set by the memory capacity of our machine. The 15 sites included were:
{\bf [000], [020], [110], [200], [220], [130], [310], [330], [420], [1-10], [-110], [240], [130], [040], [150] }, in that order.
\\
The results obtained for the energies ($E_0, E_1$) of the lowest two states relative to $N\epsilon_{s,0}$ are summarized in the following table. For the larger plaquettes, an energy cut-off was applied in order to limit the size of the matrix.

\begin{tabular}
{|c|c|c|c|c|c|c|}
\hline
Sites & $p$-states & Cut-off (meV) & States  & $E_0$ (meV) & $E_1$ (meV) & $E_0$ / Site (meV)
\\
\hline
5 & 1 &-& 80 & -2 & 26 & 0
\\
5 & 2 &-& 80 & 31 & 43 & 6
\\
6 & 1 &-& 192 & -4& 19& -1
\\
6 & 2 &-& 240 & 12 & 37 & 2
\\
8 & 1 &-& 1024 &-15 &12 & -2
\\
8 & 2 &-  & 1792 & -9&12 & -1 
\\
9 & 1 &-  & 2304 &-16&9 & -2
\\
9 & 2 & 300 & 1044 &-14 & 5 & -2
\\
12 & 1 & 300 & 2784 & -21 &-7 & -2
\\
12 & 2 & 300 & 3696 & -33 & -9 & -3
\\
12 & 3 & 300 & 2200 & -34 &-12 & -3
\\
12 & 4  & 300 & 4455 & -24 &-8 & -2
\\
15 & 2 & 300 & 9660 & -40 & -23& -3
\\
15 & 3 &300& 5915 & -44 & -25& -3
\\
15 & 4 &250&1365  & -26 & -8& -2
\\
\hline
\end{tabular}
\section{Conclusions}
The intrinsic complexity of this exact method and the inapplicablity of a perturbative approach have so far confounded our attempts to establish a lower bound on the absolute minimum site energy. It follows from the variational principle that inclusion of higher $|s,n\rangle$ states, as well as further increase in plaquette size, will result in even lower minimum energies. A mean-field approach is perhaps indicated, but we have as yet to find a sufficiently accurate formulation. It is nevertheless already clear from the above data that entangled states are favoured in the stoichiometric regime. The existence of a low temperature phase in which all the deuterons cohere in a mesoscopically entangled state is hence strongly indicated.
\section{Over-cancellation of Coulomb barrier}
At small interparticle distances :- $K|r_1 - r_2| << 1$, the off-diagonal elements of the type we have been considering are comparable in magnitude, but opposite in sign, to the static Coulomb pair repulsion term. It is hence reasonable so suppose
that, once coherence has been established, the height and width of the effective Coulomb barrier between neighbouring $s,p$ state pairs is reduced, with a concommitant increase in the - normally infinitesimally slow - D-D fusion rate. In order to investigate the magnitude of this effect, we solved the two-particle Hamiltonian for the two states discussed in connection with (\ref{hpair}) above. Both the static (\ref{TF}) and dynamic (\ref{hif}) interactions were included. Memory constraints limited us to a grid resolution of $0.06\AA$. In view of (\ref{sp5}) and (\ref{hpair}), the solution was constrained to be of the form:
\begin{equation}
\left(
\begin{array}{c}
\psi(x_1,y_1,z_1,x_2 - {a \over 2}, y_2 - {a \over 2}, z_2)
\\
-\psi(x_2 - {a \over 2},y_2 - {a \over 2}, z_2, y_1,x_1,z_1)
\end{array}
\right)
\end{equation}
where $\psi$ is odd in its 5th argument and even in all others.
\\
It was found that the lowest energy solution, with 
\begin{equation}
\epsilon \approx \epsilon_{s,0} + \epsilon_{p,0}  - {3 {\sqrt 2}e^2 d^2 \over a^3}
\end{equation}
- exhibited an increased probability for close encounters of the two hydrogen nucleii right down to the limit of our resolution.  At $|r_1 - r_2| = 0.06\AA$, the amplitude was enhanced by about an order of magnitude over the simple product state that pertains when interaction is neglected. This exciting result implies that the dipole-dipole attraction effectively over-cancels the Coulomb repulsion at least down to this length scale. The region of overlap was concentrated about the T-site lattice potential minima that are equidistant between the two O-sites. 
\section{Further work}
A search is currently being undertaken for other metallic lattices with high affinity for hydrogen and flat effective potentials. A multi-level grid DFT algorithm of high accuracy has been developed for this purpose.


\begin{thebibliography}{2}
\bibitem{Kur96}
G. Kurizki, A. Kofman, V.Yudson, Phys. Rev. {\bf A53} R35--R38 (1996).
\bibitem{Bro06}
J.Brown, arxiv.org/abs/cond-mat/0608292
\bibitem{Kri94}
H.Krimmel, L. Schimmele, C. Els\"asser, M. F\"ahnle, J.Phys. Condens. Matt. {\bf 6} 7679--7704 (1994).
\bibitem{Dye05}
M.Dyer,C.Zhang,A.Alavi, ChemPhysChem {\bf 6}, 1711--1715 (2005).
\bibitem{Pus84}
M.Puska, R.Nieminen, Phys. Rev. { \bf B29}, 5382--5397 (1984).
\end{thebibliography}
\end{document}